\documentclass[prb,preprint,showpacs,preprintnumbers,amsmath,amssymb,
     floatfix,superscriptaddress]{revtex4}
\newif\ifpdf\ifx\pdfoutput\undefined\pdffalse\else\pdfoutput=1\pdftrue\fi
   \newcommand{\pdfgraphics}{\ifpdf\DeclareGraphicsExtensions{.pdf,.jpg}\else\fi}
\usepackage{graphicx}
\usepackage{dcolumn}
\usepackage{bm}
\usepackage{amsmath} \usepackage{amsfonts} \usepackage{amssymb}
\newcommand{\chem}[1]{\ensuremath{\mathrm{ #1 }}}

\begin{document}
\pdfgraphics
\title{Mo$_6$S$_6$ nanowires: 
structural, mechanical and electronic properties}
\author{I. Vilfan}
\email[E-mail: ]{igor.vilfan@ijs.si}
\affiliation{J. Stefan Institute, Jamova 39, SI-1000 Ljubljana, Slovenia}
\date{\today}

\begin{abstract}
The properties of  $\chem{Mo_6S_6}$ nanowires were investigated with
\textit{ab-initio} calculations based on the density-functional theory.
The molecules build weakly coupled one-dimensional chains, like  
$\chem{Mo_6Se_6}$ and $\chem{Mo_6S_{9-x}I_x}$, and the crystals are 
strongly uniaxial in their mechanical and electronic properties.
The calculated moduli of elasticity and resilience along the chain axis are 
$c_{11} = 320$ GPa and $E_R = 0.53$ GPa, respectively.
The electronic band structure and optical conductivity indicate 
that the $\chem{Mo_6S_6}$ crystals are good quasi-one-dimensional conductors. 
The frequency-dependent complex dielectric tensor $\varepsilon$, calculated 
in the random-phase approximation, shows a strong Drude peak in  
$\varepsilon_\parallel$, i.e., for the electric field polarised parallel to 
the chain axis, and several peaks related to interband transitions.
The electron energy loss spectrum is weakly anisotropic and has a strong peak 
at the plasma frequency $\hbar \omega_p \approx 20$ eV.
The stability analysis shows that $\chem{Mo_6S_6}$ is metastable against the
formation of the layered $\chem{MoS_2}$.
\end{abstract}

\maketitle

\section{Introduction}
Quasi one-dimensional materials like nanowires or nanotubes are attracting 
considerable interest, in particular because of their potential applicability
in nanoelectronic and nanomechanical devices.
Common to all of them is very strong uniaxiality in the electronic as well as 
mechanical properties, which is related to weak van der Waals coupling between 
individual molecular chains. 
They are very elastic, the strain at the (ultimate) tensile strength exceeds that 
of steel often by more than two orders of magnitude. 
As a consequence, they show extraordinary tensile strength along the long 
molecule axis -- although the Young moduli are of the same order of magnitude. 
Perhaps even more challenging are their electronic 
properties. By tuning their bandgap or conductivity, one can build a 
series of very efficient and compact nanoscale electronic devices. 
In the past decade the research has largely concentrated on carbon-based 
nanotubes which often showed promising properties. 
Inorganic molybdenum-chalcogenide nanowires represent, as we shall see, 
a possible complementary or even alternative material with similar 
mechanical and electronic properties, but 
they are easier to prepare in clean, single-stoichiometry form.
 
As early as in 1980 Potel et al.~reported on the synthesis of
$\chem{M_2Mo_6S_6}$ (M = K, Rb, Cs).\cite{PCS80}
These compounds all crystallize in the $P6_3/m$ space group with the 
$\chem{Mo_6S_6}$ chains oriented parallel to the hexagonal $c$ axis, located 
along the rhombohedral $c$ edges and with the alkali metals in the 
$(2/3,1/3,1/4)$ and $(1/3,2/3,3/4)$ positions in the unit cell.

Molybdenum-chalcogenide crystals intercalated with Li  
(Li$_x$Mo$_6$Se$_6$, $0 < x < 8$) or with other metals were also investigated 
in the past, e.g., as solid-state electrodes for secondary (rechargeable) 
lithium batteries, as one-dimensional conductors \cite{Tarascon_85, VL99} and 
even as superconducting materials.\cite{BMPGS88} In the second example Li 
or Pb were needed to enhance the conductivity of nanowires. 
It seems that the extremely interesting mechanical properties of 
molybdenum-chalcogenide crystals were overlooked at that time. 
More recently, F.J. Ribeiro et al.\cite{RRC02} investigated the elastic
properties and the band structure of Li$_2$Mo$_6$Se$_6$ by performing the 
\textit{ab-initio} total energy calculations. 
They pointed out the extraordinary high theoretical value of the 
Young modulus, comparable to the modulus of carbon nanotubes. 

Other Mo-based nanowires, composed of Mo, chalcogens (S) and halogens (I) to
form $\chem{Mo_6 S_{9-x} I_x}$ (Mo-S-I) have been investigated 
recently.\cite{MKPG05}
They have a skeleton of Mo$_6$ octahedra, each dressed with six anions 
(S or I), and bound together into chains by three anions (again S or I). 
Specific to these materials is growing in bundles of identical chains, large 
Young moduli, very small shear moduli, and easily controlled electronic 
properties.
 
In this paper we concentrate on the $\chem{Mo_6S_6}$ (Mo-S) nanowires
without and with intercalated K atoms.
Using the ab-initio total-energy and band-structure calculations we 
predict and discuss the stability, elastic and electronic properties of 
crystals made of  $\chem{Mo_6S_6}$.  

\section{Structure and computational details}
\begin{figure} [tb]
    \begin{minipage}{0.55\columnwidth}
    \begin{center} 
   \includegraphics[width=0.85\columnwidth]{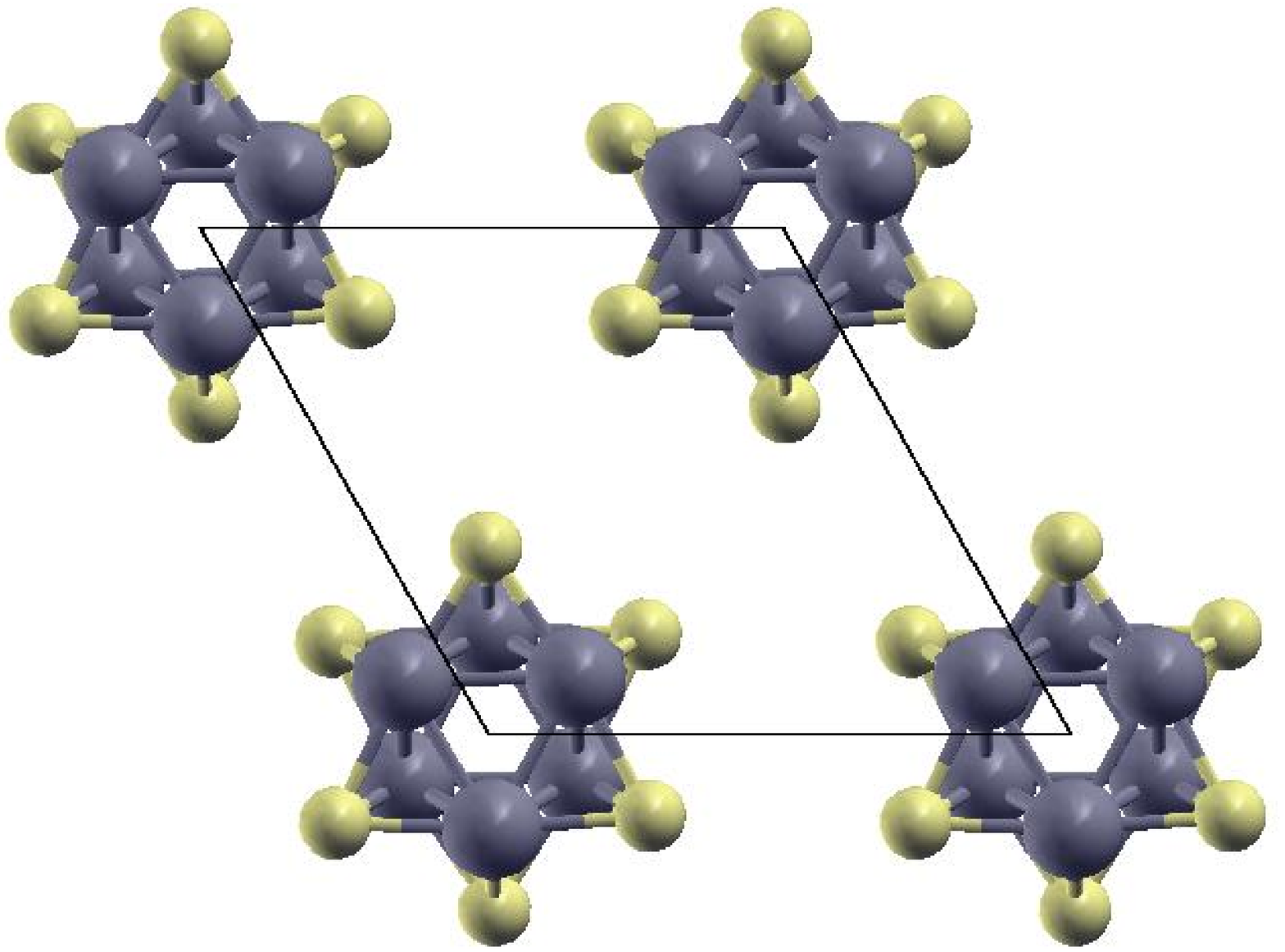}
    \end{center}
   \end{minipage}
\begin{minipage}{0.40\columnwidth}
   \begin{center} 
   \includegraphics[width=0.95\columnwidth]{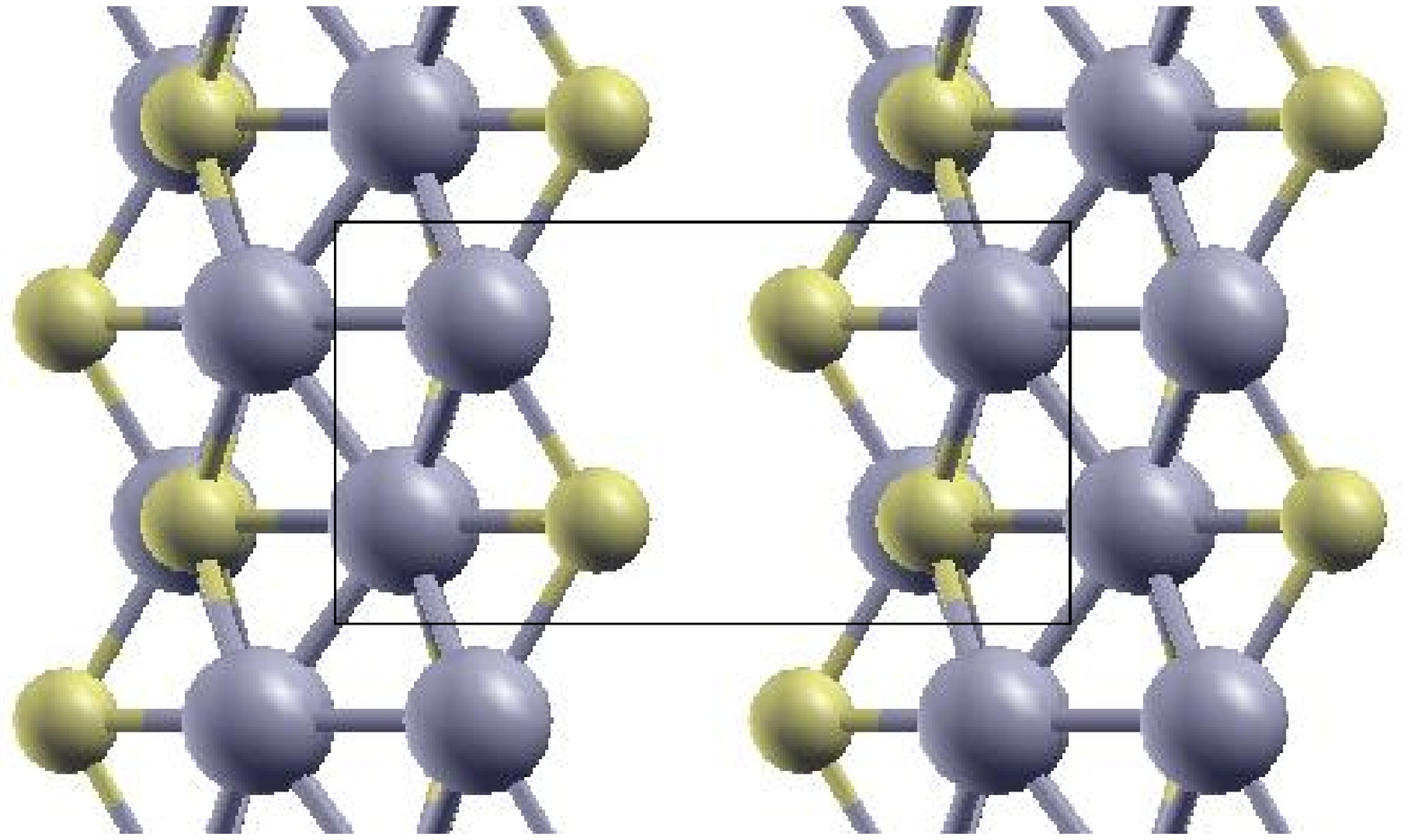}
    \end{center}
    \end{minipage}\\
\begin{minipage}{0.40\columnwidth}
   \begin{center} {\large a}
   \end{center}
   \end{minipage}
\begin{minipage}{0.40\columnwidth}
   \begin{center} {\large b}
    \end{center}
    \end{minipage}
   \caption{\label{mo6s6} 
     The crystal of $\chem{Mo_6S6}$. 
     (a) top view (along the [0001] direction) and (b) side view (along the 
     [1210] direction).
     Grey (dark) spheres: Mo, yellow (light) spheres: S.
     The Mo$_6$S$_6$ molecules form chains along the [0001] direction
     and a bundle of chains forms a nanowire.
     The separation between the chains is large and the interaction between
     them is weak, the Mo$_6$S$_6$ crystal is strongly anisotropic.
     In  $\chem{K_2Mo_6S6}$ the $\chem{Mo_6S6}$ chains are rotated by 
     $\sim 20^\circ$ in the anticlockwise direction and the K atoms are
     in the $(2/3,1/3,1/4)$ and $(1/3,2/3,3/4)$ positions.}
\end{figure}

The crystal structure of Mo-S is shown in figure~\ref{mo6s6}. 
The hexagonal crystals comprise one-dimensional chains oriented along the 
hexagonal $c$ axis. Each chain is formed  of a stack of Mo$_3$ triangles
with staggered rotations and dressed with three anions (S). 
In order to clarify the stability, possible structures and properties of Mo-S 
nanowires we performed extensive density-functional theory (DFT) 
analysis  using the WIEN2k code.\cite{BSMKL01} 
The simulations were performed on a mixed basis set of augmented plane waves plus
local orbitals (APW+lo) for low orbital momenta ($\ell \le 2$) and linearised 
augmented plane waves (LAPW) for all the higher orbital momenta.\cite{SNS00} 
The exchange and correlation potential was treated in the 
Perdew, Burke and Ernzerhof generalised-gradient approximation (GGA).\cite{PBE96} 
The muffin-tin radii were set to 
$R_{\rm Mo}^{\rm MT} = 1.06$ {\AA} for Mo,        
$R_{\rm S}^{\rm MT}  = 1.27$ {\AA} for S and      
$R_{\rm K}^{\rm MT}  = 1.32$ {\AA} for K,         
the kinetic energy cutoff was 
$E_{\rm max}^{\rm wf} = 12.3$ Ry and the plane-wave expansion cutoff was 
$E_{\rm max}^{\rm pw} = 196 $ Ry.          
The total energy was calculated on a 
mesh of 150 ($9 \times 9 \times 18$) $k$-points in the irreducible part of the 
Brillouin zone. The problem of the above DFT method is inappropriate 
treatment of the weak long-range van der Waals interaction which is 
responsible for bonds between individual chains in such molecular crystals.
As a consequence, the DFT treatment of interaction between individual chains 
is not very reliable. 
On the other hand, the analysis of $\chem{K_2Mo_6S_6}$ chains reveals that the 
calculated hexagonal lateral lattice 
constant $a$ (at $T=0$ K, as usual in the DFT calculations) differs
from the experimental one (at room T) only by 2.6\%, see Table \ref{latt}.
In any case, it is clear from both theory and experiment, 
that the lateral interaction between the chains is very weak. 
This is another reason why we shall concentrate on the intrachain 
structure and properties of single chains.

In the next stage, the total energy was minimised with respect to the
internal atomic coordinates and to the hexagonal lattice constants $a$ and $c$
by assuming that $\chem{Mo_6S_6}$ belongs to the same space group as
$\chem{K_2Mo_6S_6}$, i.e.,  $P6_3/m$ (176).\cite{PCS80}
This assumption is corroborated by weak coupling between individual molecular 
chains and on extensive analysis of Mo-S-I 
nanowires.\cite{MKPG05} The space group $P6_3/m$ allows rotational 
relaxation of molecular chains around the $c$ axis but does not allow mutual 
translations or deformations of chains along the $c$ axis. 
The results of the energy minimizations 
of $\chem{Mo_6S_6}$ and $\chem{K_2Mo_6S_6}$ are compared with 
the available experimental lattice constants in Table \ref{latt}.
The structure of  $\chem{Mo_6S_6}$ is shown
in figure \ref{mo6s6} and the interatomic distances in Table \ref{dist}. 
In  $\chem{K_2Mo_6S_6}$ the calculated lattice constant $a_\textrm{calc}$ 
is $ 2.6$ \% larger than $a_\textrm{exp}$, most probably because the DFT GGA 
overestimates the dispersive forces, acting between the chains. 
The  nearest neighbouring distance between K atoms in  $\chem{K_2Mo_6S_6}$ 
($d_{NN} = c$) is close to the interatomic distances in metallic K ($d= 4.53$ {\AA}).
 
The interatomic distances between the atoms of the same $\chem{Mo_6S_6}$ 
chain are up to 8\% longer than the intrachain distances reported for 
$\chem{Mo_6S_3I_6}$ chains.\cite{MKPG05} 
The in-plane Mo-Mo distance agrees, within the precision of our DFT 
calculation, with the experimental value for the room-temperature 
bulk metallic Mo (2.73 \AA). 
In the Supplement of Ref.~\onlinecite{MKPG05}
an agreement within 0.1 {\AA} between the room-temperature 
experimental distances in $\chem{Mo_6S_3I_6}$ and the $T=0$ distances 
calculated by the DFT was reported.

\section{Results and discussion}
To investigate the stability of the $\chem{Mo_6S_6}$ and  $\chem{K_2Mo_6S_6}$ 
nanowires we performed 
also the DFT simulation of the layered MoS$_2$ compound and of atomic S and K
in vacuum (i.e., one S or K atom in the unit cell $10.1\times 10.6\times 11.1$ 
\AA$^3$ with periodic boundary conditions).  
The comparison shows that for  $\chem{Mo_6S_6}$ the
energy difference with respect to the more stable $\chem{MoS_2}$ is 
$\Delta E = E(\textrm{Mo$_6$S$_6$}) - 6 E(\textrm{MoS$_2$}) + 6 E(\textrm{S})
= 28$ eV (per one Mo$_6$S$_6$ unit) and tells us that the Mo-S nanowires 
with Mo in oxidation state 2 are, like other Mo-S-I nanowires, metastable 
against MoS$_2$ with the Mo oxidation state 4. 
The stability of $\chem{Mo_6S_6}$ is comparable to the stability of 
already synthesised material $\chem{Mo_6S_3I_6}$ for which the calculated
energy difference is 
$\Delta E = E(\chem{Mo_6S_3I_6}) - 6 E(\chem{MoS_2}) + 9 E(\chem{S})
-6 E(\chem{I}) = 32$ eV. Per one atom, however, the energy difference is 
2.33 eV for $\chem{Mo_6S_6}$  and 2.13 eV for  $\chem{Mo_6S_3I_6}$. 
Therefore we expect  $\chem{Mo_6S_6}$ to be stable in the air, in analogy with 
the Mo-S-I nanowires.
The already synthesized $\chem{K_2Mo_6S_6}$ is stable in the air,\cite{PCS80}
the corresponding energy difference to $\chem{MoS_2}$ is 
21.0 eV and the Mo oxidation state 2.33.
\subsection{Elastic properties}
\begin{figure}
   \includegraphics[width=0.8\columnwidth]{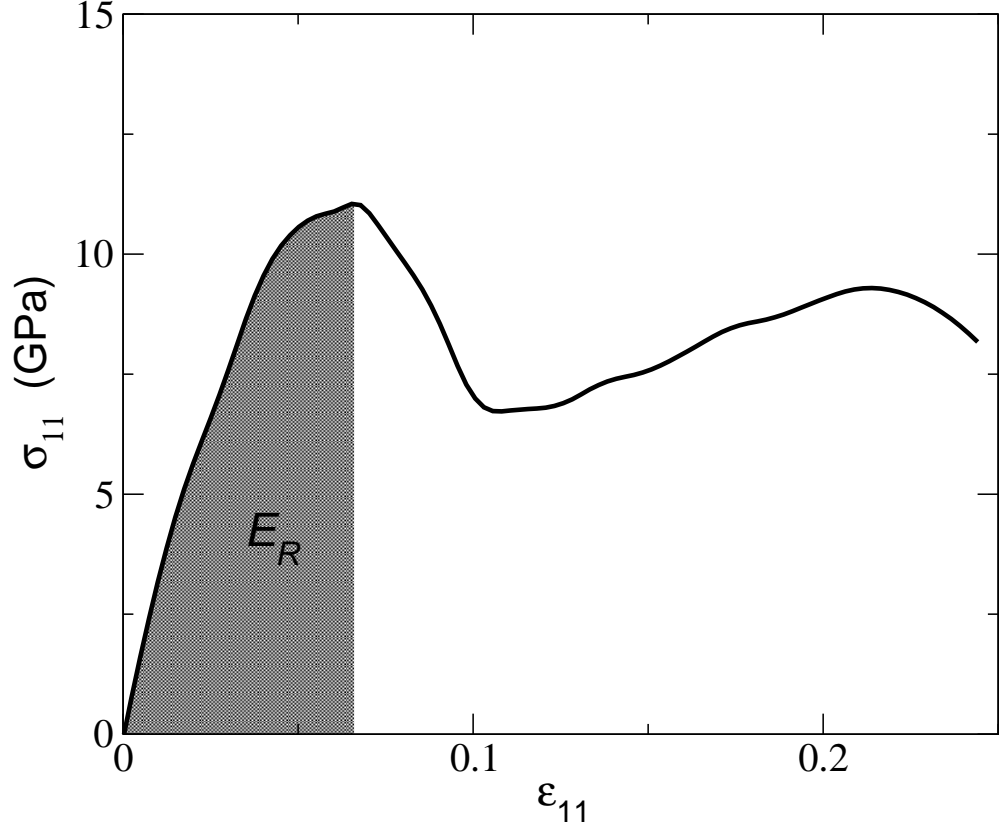}
   \caption{\label{young}
     Elastic properties of Mo$_6$S$_6$ nanowires. 
     The elastic strain $\epsilon_{11}$ is for a unit cell,
     doubled along the $c$ axis.
     At the first and second maxima the Mo-Mo bonds and
     the Mo-S-Mo bonds break, respectively. 
     The initial slope of the curve defines the elastic modulus
     $c_{11} = 320$ GPA,
     and the grey area the modulus of resilience $E_R = 0.53$ GJ/m$^3$.}
\end{figure}

\begin{figure}
\begin{minipage}{5.0cm}
   \includegraphics[width=3.0cm]{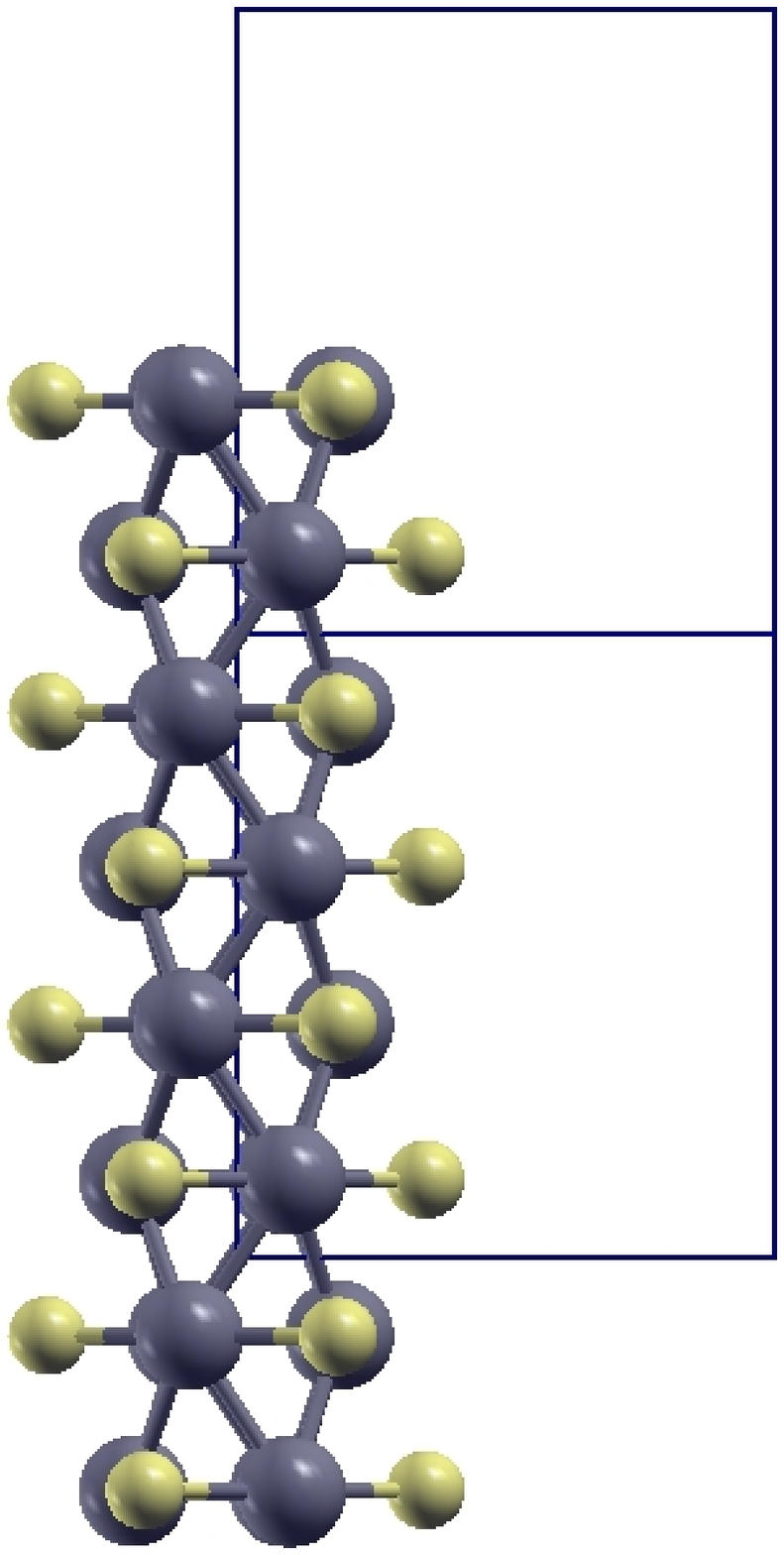} 
\end{minipage}
\begin{minipage}{5.0cm}   
   \includegraphics[width=3.0cm]{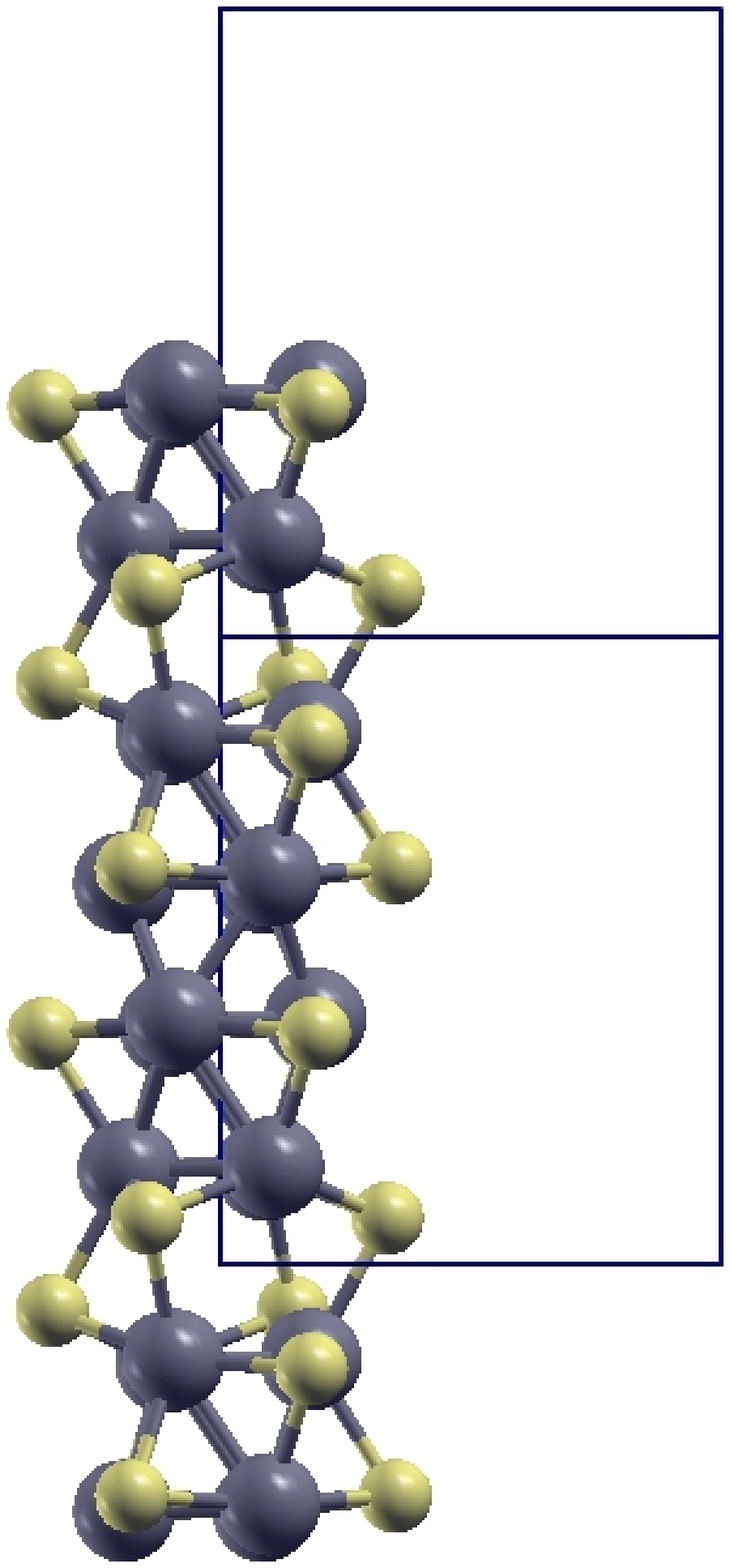}
\end{minipage}
\begin{minipage}{5.0cm}
   \includegraphics[width=3.0cm]{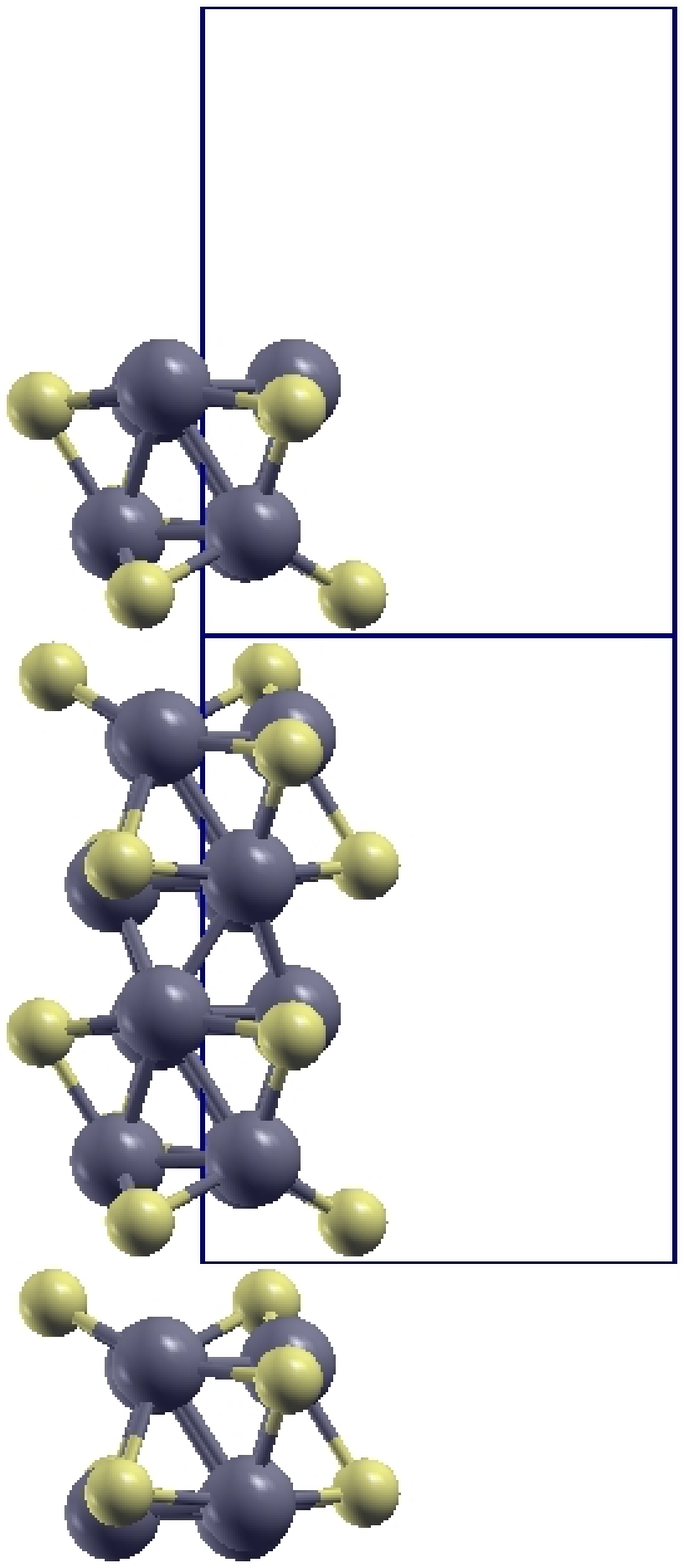}
\end{minipage}\\
\begin{minipage}{5.0cm}
   \begin{center} {\large a}
   \end{center}
   \end{minipage}
\begin{minipage}{5.0cm}
   \begin{center} {\large b}
    \end{center}
    \end{minipage}   
\begin{minipage}{5.0cm}
   \begin{center} {\large c}
    \end{center}
    \end{minipage}   
\caption{\label{exp}
    Stretching of Mo$_6$S$_6$ nanowires is a two-stage process, as shown 
    on this graph with two unit cells in the direction of the chain.
    (a) Fully elastic regime before the Mo-Mo bonds at a particular chain 
    segment start to tear apart ($c = 4.60$ \AA).
    (b) The Mo-Mo bonds between two neighbouring unit cells start to 
    break and the S atoms move into the empty space between them
    by keeping the two ends of the broken chain together via  four Mo-S-Mo
    bonds ($c = 4.975$ \AA).
    (c) Eventually also the Mo-S-Mo bonds break ($c = 5.29$ \AA).
    }
\end{figure}

The Mo-based nanowires and nanotubes have large elastic moduli in the
direction of the wires and very small shear moduli in general.\cite{KMRM03} 
To investigate the mechanical properties of the Mo-S nanowires, we 
performed DFT simulations for a series of lattice constants $c$. 
Due to weak interchain coupling the effect of $a$ on the 
total energy is very small and was neglected  unless otherwise stated.
The lattice constant $a$ was thus kept fixed during expansion, but the 
atomic positions were fully relaxed. 
For larger lattice constants, when the nanowires start to break, we 
simulated double cells, composed of two unit cells in the $c$ direction,
so that each simulated cell contained 12 Mo and 12 S atoms
and the space group was reduced to $P\overline{1}$.
In this way we reduced the effect of the interaction between breaking 
segments in a simulation with periodic boundary conditions.
The results, presented in figure~\ref{young} show the behaviour of 
the stress $\sigma_{11}$ ($\sigma_{11} \propto \textrm{d}E/\textrm{d}c$ at constant 
$a$) under tensile strain $\epsilon_{11}$. 
The initial slope yields for the elastic coefficient in the direction of the 
nanowire $c_{11} \approx 320$ GPa, which is larger than the Young modulus of steel 
but about three times smaller than for (single-wall) carbon nanotubes, see
Table \ref{young_table}. 
Upon increasing strain, the slope decreases and eventually changes sign.
Usually we say that the system breaks after $\sigma_{11}$ reaches the maximum
stress. Mo-S, however, behaves in a very different way. $\sigma_{11}$ decreases
after the first maximum, but then it starts to rise again until it
reaches a local maximum. 
Breaking of Mo-S nanowires is thus a two-stage process, see fig.~\ref{exp}. 
In the first stage, the Mo-Mo bonds at some place are stretched 
until they break. 
In the second stage, the nearby S atoms move into the empty space between
the chains and bind together the two ends of the wires until, eventually,
also these Mo-S-Mo bonds and thus the whole chain breaks at the second 
maximum in figure~\ref{young}.
Notice that the nonlinear regime below $\epsilon_{11} \sim 0.6$ in fig.~\ref{young} 
is the consequence of the anharmonic contributions to the interatomic potential 
and is not related to plastic deformation which is absent in our DFT simulations.  

The modulus of resilience, i.e., the  maximal energy that can be 
\textit{reversibly} stored in an elastic medium, is in our case equal to 
\begin{equation}
   E_R = \int_0^{\epsilon_{11}^\textrm{max}} 
         \sigma_{11} \textrm{d}\epsilon_{11}.
\end{equation}
The Mo-S nanowires have a very high modulus of resilience, 
$E_R \approx 0.53$ GPa (see Table \ref{young_table}).
This value is more than two orders of magnitude higher than for steel 
and already indicates possible applications of these nanowires: nanosprings.
The \textit{ideal} tensile strength, i.e., the maximal stress is 
$T \approx 11$ GPa, it is extremely high for an inorganic nanowire, but is by
a factor of 3 smaller than for carbon nanotubes. 
Stretching the wire along the $c$ axis by 6.4\% causes a 3.7\%  elongation 
of the nearest interplanar Mo-Mo bond length and only a 2.6\% contraction of the
in-plane Mo-Mo bond length, see Table \ref{dist}. 
The wire breaks at $\epsilon_{11} = 6.6$~\%. 

These are of course theoretical values. The experimental values
of the tensile strength will be smaller due to lattice imperfections. 
Due to the almost one-dimensional structure and consequently small shear modulus, 
broken chains will diminish the tensile strength of an entire bundle of chains.

\subsection{Electronic properties}
\begin{figure}
   \includegraphics[scale=0.3]{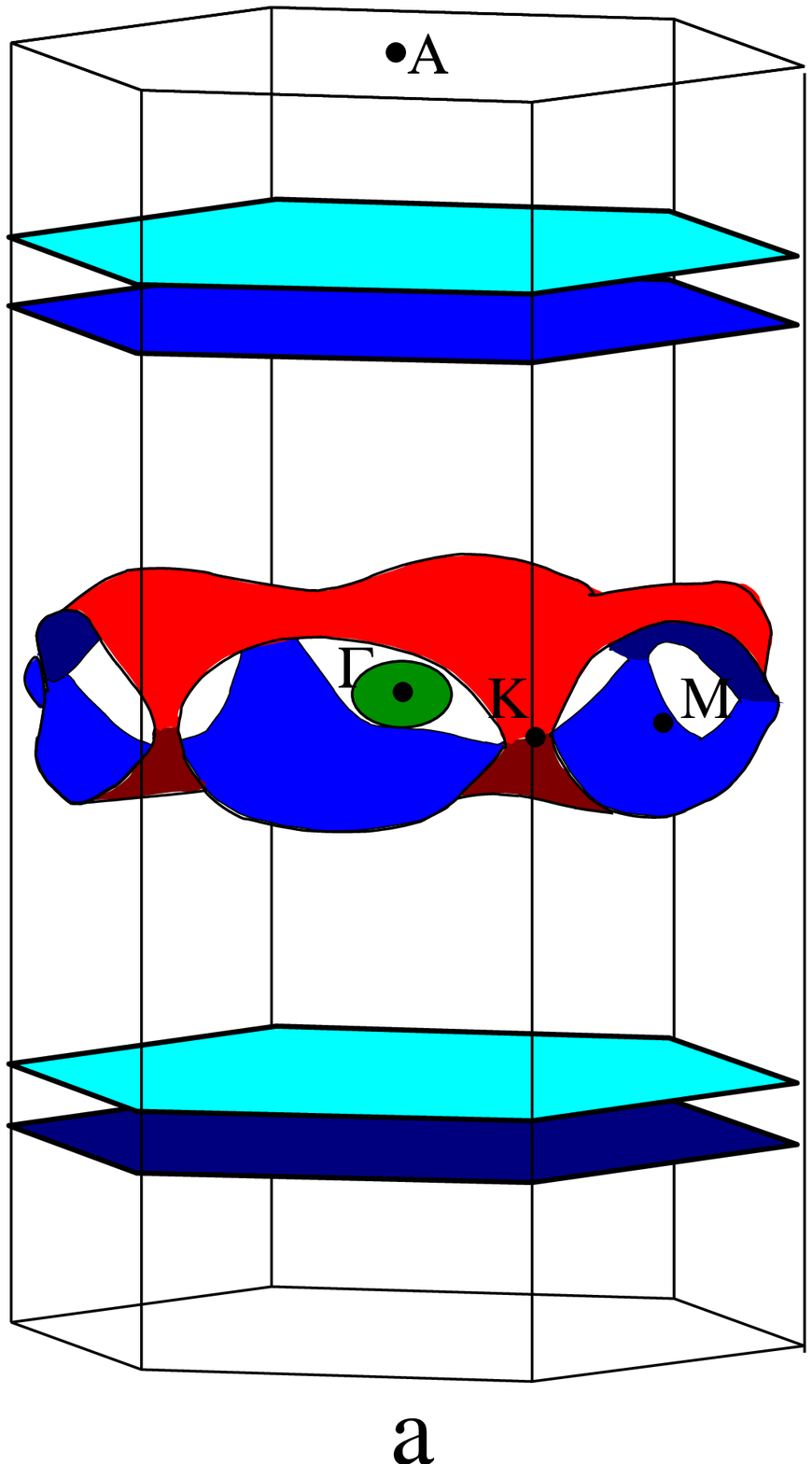}\\
   \includegraphics[scale=0.5]{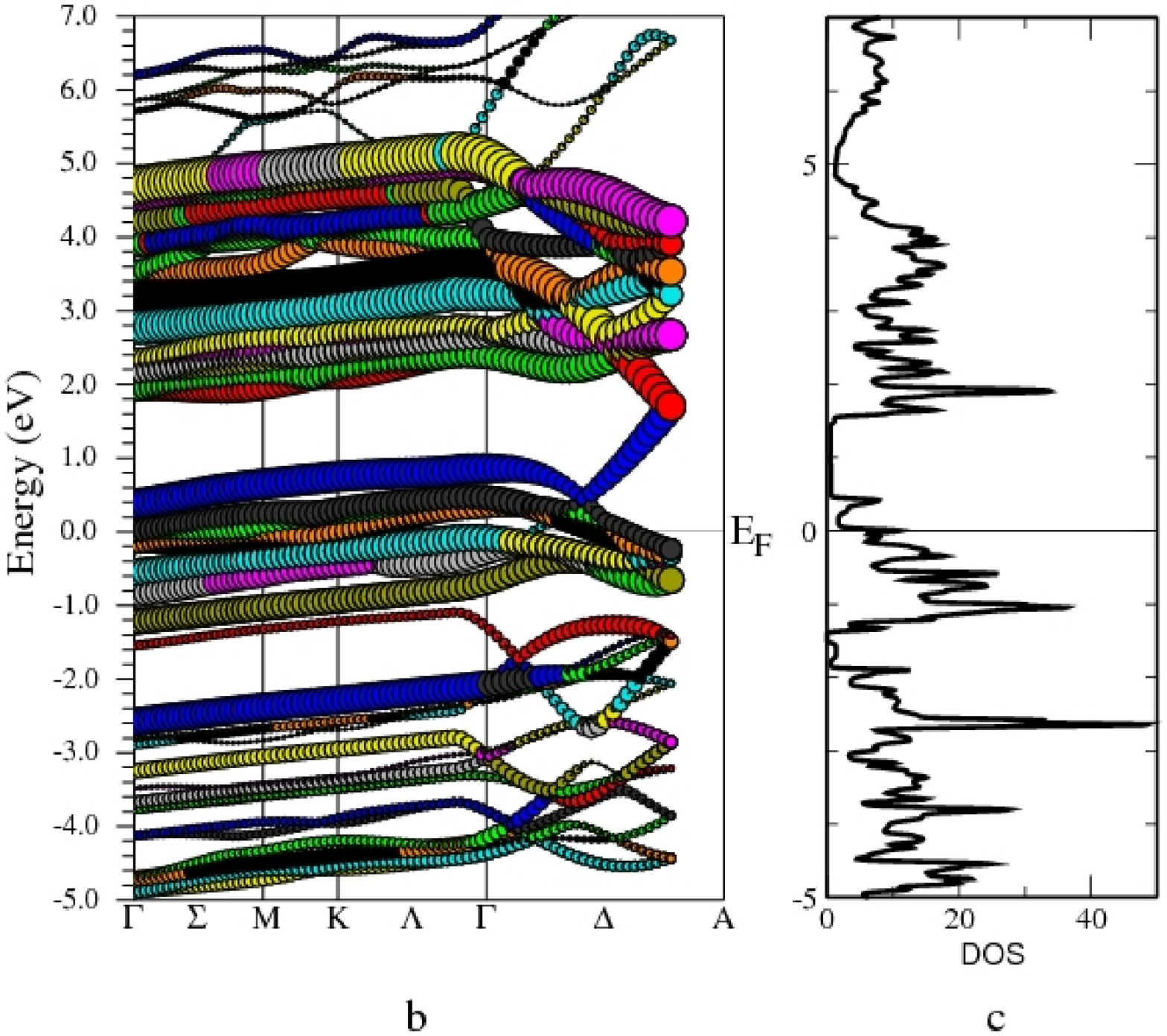}
   \caption{\label{bands}
   (a)  The Fermi surfaces in the Brillouin zone of $\chem{Mo_6S_6}$ are very 
   anisotropic and reflect the fact that the coupling between the molecular 
   chains is weak. As a consequence, the electron group velocity points 
   mainly parallel to the $c$ axis. The letters show the positions of 
   symmetry points in the Brillouin zone.
   (b) Electron band structure along the main symmetry directions. 
   The dispersion
   of occupied states in the lateral direction is small, reflecting weak
   coupling between the chains. 
   The radii of the circles indicate the Mo$-4d$ character 
   of the bands. 
   The main contribution to the static charge-carrier transport comes from two 
   bands, crossing the Fermi energy close 
   to the $\Delta$ point. 
   (c) The total electron density of states per eV in a unit cell.
   }
\end{figure}

\begin{figure}
   \includegraphics[scale=0.5]{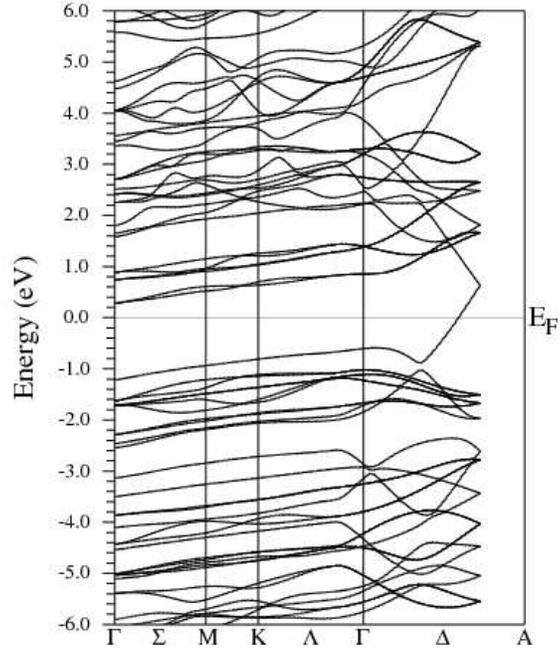}
   \caption{\label{k2bands} 
   Electron band structure of $\chem{K_2Mo_6S_6}$ along the main symmetry directions. 
   The main difference to the bandstructure of $\chem{Mo_6S_6}$ is the shift in $E_F$.}
\end{figure}

The basis of electronic and optical properties of $\chem{Mo_6S_6}$ is the 
electron band structure, shown in figure~\ref{bands}. 
For clarity, let us assign the sub-bands below -1.4 eV to the valence band 
and all the higher electron states to the conduction band. 
The Fermi energy then lies in the conduction band. 
Some of the sub-bands are doubly degenerated so that the conduction band
carries 20 electrons per unit cell in total and all the valence bands together
another 52 electrons. These numbers agree with the total number of valence
electrons in the unit cell: there are 12 atoms in the unit cell, each carrying
6 valence electrons.
Due to large  number of equivalent atoms in the unit cell
it is not possible to uniquely assign particular sub-bands to individual
atoms. Some general observations can nevertheless be made.
The band at -12.5 eV (not shown in figure~\ref{bands}) is clearly the $3s-$band 
of S atoms. Between -5 and 0 eV we have a mixture of Mo-$4d$ and S-$3p$ bands,
between 0 and 5 eV the bands have a clear Mo-$4d$ character, whereas above 5 eV
the S-$3d$ states start to prevail. In the whole energy interval there is also 
a substantial contribution of interstitial electrons to the total density of
states.   
For comparison, the band structure of $\chem{K_2Mo_6S_6}$ is shown in 
fig.~\ref{k2bands}.
The intercalation of K atoms does not alter the general shape and anisotropy 
of the bands which is still governed by the intrachain wavefunctions overlap. 
K atoms add two electrons per unit cell which occupy one band and shift the 
Fermi energy up by about 1.5 eV.
The above band structures are also similar to the calculated band structures of 
$\chem{Mo_6Se_6}$ and  $\chem{Li_2Mo_6Se_6}$ nanowires.\cite{RRC02}

Dispersion of individual occupied sub-bands in the plane 
perpendicular to the wires is very small, of the order 0.2 eV and is similar 
in magnitude to the dispersion in Mo-Se and other Mo-S-I nanowires.\cite{RRC02,MKPG05}
The reason for small dispersion lies in weak, mainly Van der Waals coupling 
between the wires. 
Interestingly enough, even intercalation with K does not alter the dispersion in 
this plane noticeably.
The conduction electrons are localised on individual wires and the 
hopping rates between the neighbouring wires are small.
The dispersion in the direction parallel to the wires is more system-specific. 
In $\chem{Mo_6S_6}$ there are four sub-bands crossing the Fermi
energy $E_F$, thus building four Fermi surfaces, shown in figure 
\ref{bands}(a). 
The first surface is a small, weakly anisotropic 
sphere close to the $\Gamma$ point which makes very small contribution to the 
charge density at  $E_F$ and to the conductivity.
The main contribution to the static charge carrier transport and to static 
dielectric properties comes from electrons at the upper two Fermi surfaces 
which cross the $\Gamma - A$ line close to the $\Delta$ point and are  
very flat as a consequence of anisotropic coupling between the wires.
Thus, electrons in two sub-bands  
contribute to the static charge transport, the other either occupy sub-bands 
that don't cross the Fermi energy or have high effective mass.
These two sub-bands have the dispersion in the direction parallel to the 
wires typical for nearly-free electrons. 
We find one electron sub-band starting at $E(\Gamma) = -0.73$ eV with an effective 
mass $m_\parallel^\textrm{el} \sim 1.2 m$ ($m$ is the free-electron mass) 
and one hole sub-band starting at $E(\Gamma) = 0.37$ eV with an effective mass 
$m_\parallel^\textrm{h} \sim 0.8 m$.
The slopes at $E_F$ of these two sub-bands give the electron group velocity 
$v_\parallel^\textrm{el} \sim 0.4 \times 10^6$ m/s and the hole group 
velocity $v_\parallel^\textrm{h} \sim 0.6 \times 10^6$ m/s.
The spin-orbit coupling -- if added --  neither splits the two subbands 
that cross the Fermi energy close to  $\Delta$ nor does it alter 
noticeably the effective masses or the velocities of the two sub-bands. 
The values of $m_\parallel$ and  $v_\parallel$ are of the orders of 
magnitude typical for metals, the Mo-S  nanowires are thus candidates for 
one-dimensional ballistic quantum wires where the resistance is determined 
by the contacts of the wire. The ballistic regime is limited by the ability to 
produce  defect-free nanowires, i.e., the length of the wire must not
exceed the scattering length of the electrons. 
For a lifetime broadening of 0.01 eV (lifetime $\tau \sim 7 \times 10^{-14}$ s)
this condition limits the length of the wires to $\sim 40$ nm.  
As one-dimensional conductors, the Mo-S nanowires are also candidates for 
a Peierls transition to an insulator at low temperatures. 
However, we cannot give an estimate here because we did not investigate
the electron-phonon interactions. 

Let us compare the electron bands in $\chem{Mo_6S_6}$ with those in 
$\chem{Mo_6S_9}$ (three additional S atoms are ordered in triangles in the 
bridging planes between successive $\chem{Mo_6}$ octahedra, see 
Ref.~\onlinecite{MKPG05}) and in $\chem{Mo_6S_6I_3}$ (in this case the Mo 
octahedra are ``intercalated'' with 3 I atoms instead of 3 S).
In the latter two cases the sub-band dispersion in the $z$ direction is smaller 
and the band structures have narrow gaps close but above the Fermi energy.
The Mo-$4d$ electrons in $\chem{Mo_6S_6}$ build conduction 
channels along the chains which are interrupted in case of $\chem{Mo_6S_9}$
and $\chem{Mo_6S_6I_3}$. As a consequence, the $\chem{Mo_6S_6}$ is expected
to be a much better conductor than Mo-S-I chains of the type  
$\chem{Mo_6X_9}$ (X is a combination of S and I).

\subsection{Optical properties}
During synthesis of Mo-S-I nanowires several different stoichiometries are
obtained which then have to be purified in order to obtain a 
single-stoichiometry compound. 
Therefore it is essential to distinguish between them.
To facilitate the  characterisation of different stoichiometries we
predict here also the optical properties of $\chem{Mo_6 S_6}$.

\begin{figure}
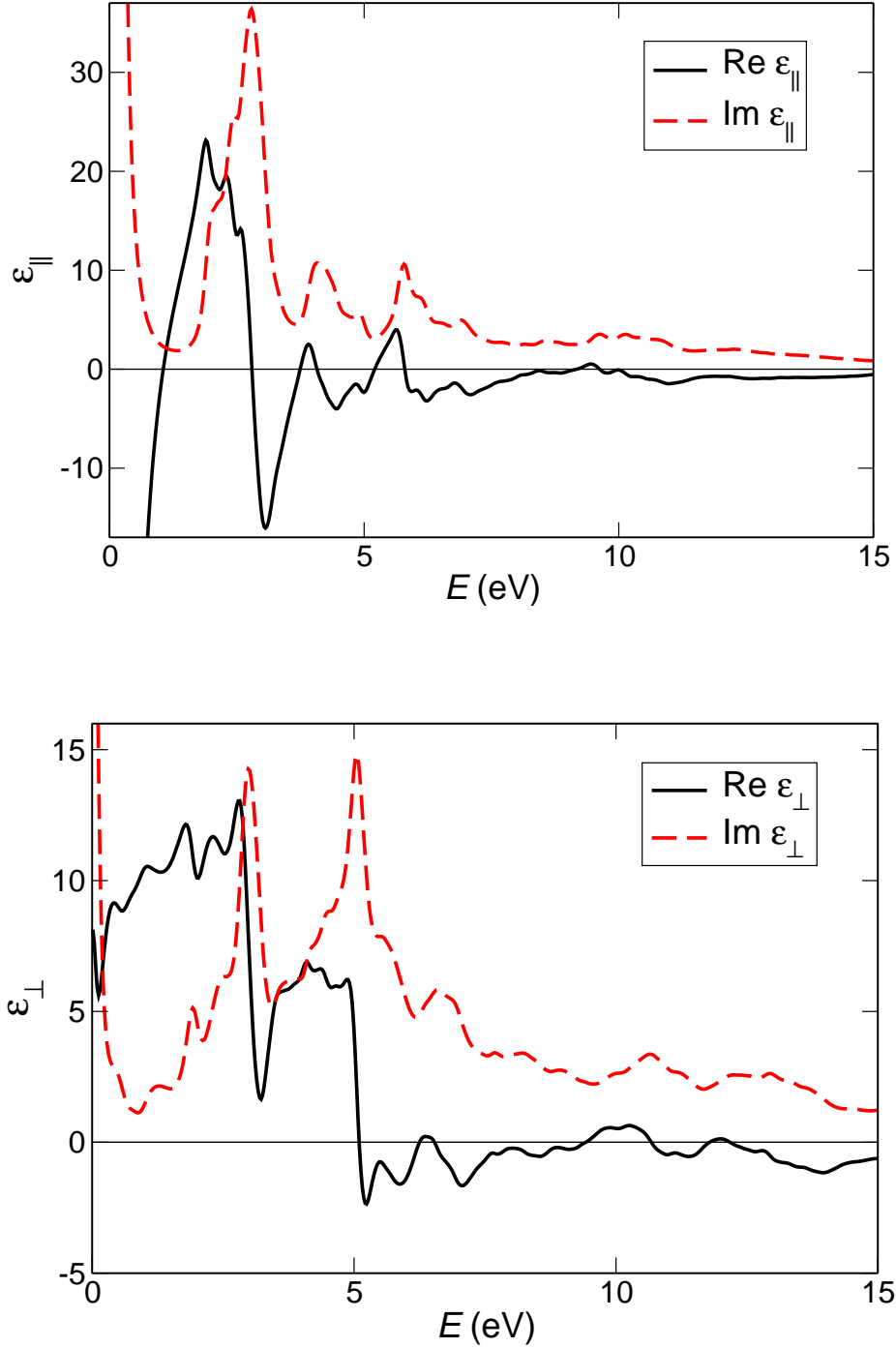

   \includegraphics[scale=0.5]{Figs/mo6s6_epsilon_zz}\vskip1.5cm
   \includegraphics[scale=0.5]{Figs/mo6s6_epsilon_xx}
   \caption{\label{eps}
   Energy dependence of the complex dielectric functions 
   $\varepsilon_{\parallel}$ and  $\varepsilon_{\perp}$,  broadened by a 
   Lorentzian with $\Gamma = 0.1$ eV.  
   The dielectric function $\varepsilon_{\parallel}$ has a pronounced 
   Drude peak, typical for intraband transitions in metals,
   and an absorption peak at 2.8 eV.
   $\varepsilon_{\perp}$ shows two pronounced absorption peaks which are 
   associated to the interband transitions and are closely related to the 
   maxima in the joint density of states, which means that most
   of the sub-bands in the valence and conduction bands are a mixture of
   several symmetries, they have no unique band character.
   As a consequence, the selection rules are not very selective.
   }
\end{figure}

We start with the frequency dependent complex dielectric function,  
figure~\ref{eps}. 
The imaginary part of the dielectric tensor ${\varepsilon}$ in the limit 
$q \to 0$ is calculated  with the WIEN2k code  on a tetrahedral mesh of 300  
$k-$points in the irreducible part of the Brillouin zone and in the 
random-phase approximation.\cite{CAD}
The real part is then obtained from $\textrm{Im}(\varepsilon)$ with the 
Kramers-Kronig relations. 
In our case the $\varepsilon$ tensor is diagonal with the components 
$\varepsilon_{zz} = \varepsilon_\parallel$ and $\varepsilon_{xx} = 
\varepsilon_{yy} = \varepsilon_\perp$. 
The dielectric tensor has a Drude peak, associated with the transitions 
between the conduction electrons close to $E_F$, which is more pronounced in 
$\varepsilon_\parallel$. 
The anisotropy in ${\varepsilon}(E\to 0)$ is a direct consequence of the 
shape of Fermi surfaces. Above the Drude peak, $\varepsilon_\perp$ has two 
pronounced absorption peaks at 3 and 5 eV whereas  
$\varepsilon_\parallel$ has a structured absorption peak between 2 and 2.8 eV. 
These peaks are related predominantly to interband transitions between
different S$-3p$ (below $E_F$) and Mo$-4d$ (above $E_F$) sub-bands.
Above 5 eV the transitions are mainly to/from the S$-3d$ sub-bands.
In all the cases, however, the influence of interstitial electrons is large. 
Coming from the high-frequency side, $\textrm{Re}(\varepsilon)$ 
changes sign (vanishes) at the plasma frequencies $\hbar \omega_{p\perp} = 20.1$ eV
and $\hbar \omega_{p\parallel} = 20.0$ eV. 
The plasmons are \textit{longitudinal collective excitations} 
of \textit{all} valence and conduction electrons.
$\omega_p$ agrees well with the estimate $\omega_p^2 = n e^2/\varepsilon_0 m$
($e$ is the electron charge, $m$ its mass and  $\varepsilon_0$ the  permittivity of 
free space). With $n = 72$ valence electrons per unit cell of the volume 
$V = 317.6$ {\AA}$^3$ we get $\hbar \omega_p = 17.6$ eV. 
The plasma frequency can be also tested with the $f$-sum rule,
\begin{equation}
   \frac{\pi}{2} \omega_{p,i}^2 = \int_0^\infty \textrm{Im} 
   [\varepsilon_{ii}(\omega)]\omega\textrm{d}\omega.
\end{equation}
Numerical integration in the range from 0 to 41 eV gives 
$\hbar \omega_{\perp} = 18.5$ eV and $\hbar \omega_{\parallel} = 18.9$ eV.
At the plasma frequency the effect of the anisotropy in the band structure on 
the dielectric tensor is small and $\varepsilon$ is almost isotropic.

\begin{figure}
   \includegraphics[scale=0.5]{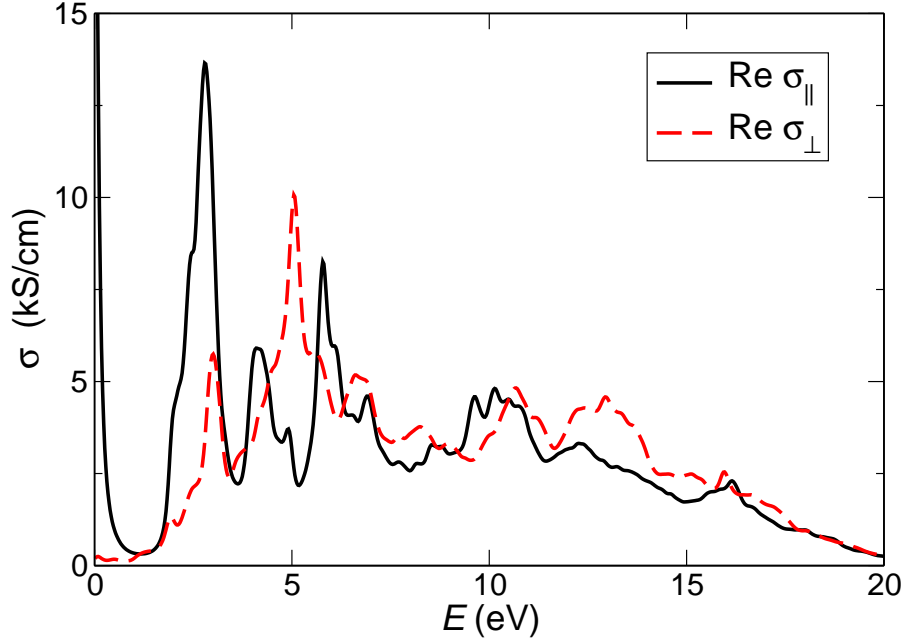}
   \caption{\label{sigma}
    Spectral dependence of the optical conductivity components 
    $\textrm{Re}(\sigma_{\parallel})$ and  $\textrm{Re}(\sigma_{\perp})$.  
    Clearly seen is strong anisotropy in the static conductivity. 
    The otherwise diverging peak at $E=0$ was damped with $\Gamma_\parallel
    = \Gamma_\perp = 0.1$ eV to mimic the charge-carrier scattering.
    }
\end{figure}

From $\varepsilon$ we get the optical conductivity $\textrm{Re}[\sigma(\omega)]
=\varepsilon_0 \omega \textrm{Im}[\varepsilon(\omega)]$, see figure~\ref{sigma}.
The static ($\omega\to 0$) peak is lifetime broadened due to the charge-carrier
scattering mainly on impurities and imperfections in the crystal lattice.
In the figure, the Drude peaks are broadened with 
$\Gamma_\parallel = \Gamma_\perp = 0.1$ eV, which corresponds to a lifetime
$\tau \approx 0.7\times 10^{-14}$ s. 
With such $\Gamma$ the static conductivity along the wires is 
$\sigma_\parallel \sim 2\times 10^4$ S/cm and the predicted anisotropy in
$\sigma$ is $100 : 1$.
At higher frequencies ($\hbar\omega > 5$ eV) the anisotropy in ${\varepsilon}$ 
gradually disappears since the electrons are excited also above the 
ionisation threshold of $4.36$ eV into the unbound (free) electron states.

\begin{figure}
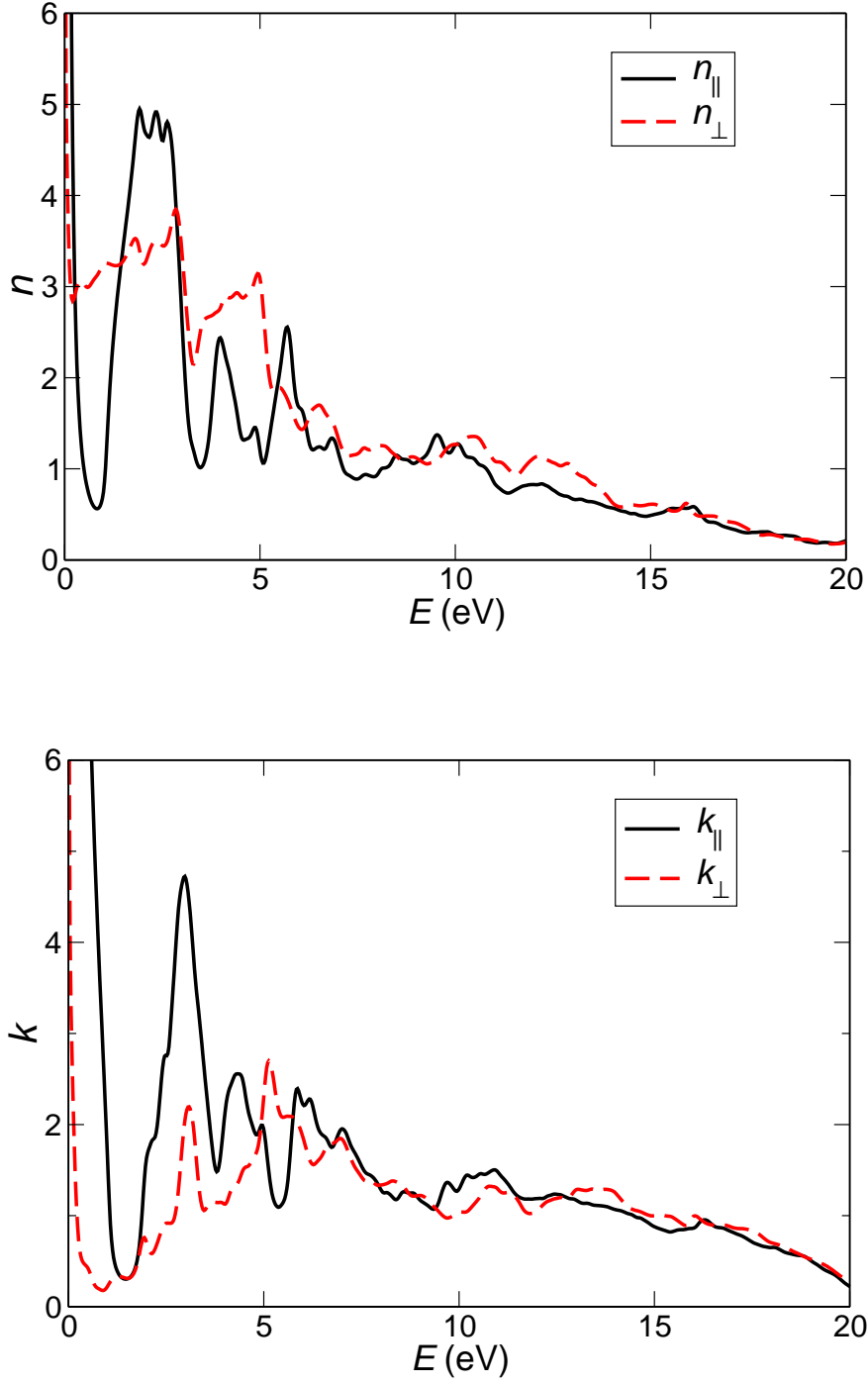

   \includegraphics[scale=0.5]{Figs/mo6s6_n}\vskip1.5cm
   \includegraphics[scale=0.5]{Figs/mo6s6_k}
   \caption{\label{refrac}
    Complex index of refraction $(n + \textrm{i} k)$, consisting of 
    $n_\parallel$ and $n_\perp$ (upper panel) and extinction 
    coefficients $k_\parallel$ and $k_\perp$ (lower panel).
    }
\end{figure}

The complex index of refraction $(n + \textrm{i} k)$ with the components
\begin{equation}
n_{ii} = \sqrt{\frac{|\varepsilon_{ii}| + 
\textrm{Re}(\varepsilon_{ii})}{2}}
\end{equation}
and
\begin{equation}
k_{ii} = \sqrt{\frac{|\varepsilon_{ii}| - 
\textrm{Re}(\varepsilon_{ii})}{2}}
\end{equation}
is shown in figure~\ref{refrac}. It has similar trends as the dielectric function 
from which it is derived. 

\begin{figure}
   \includegraphics[scale=0.5]{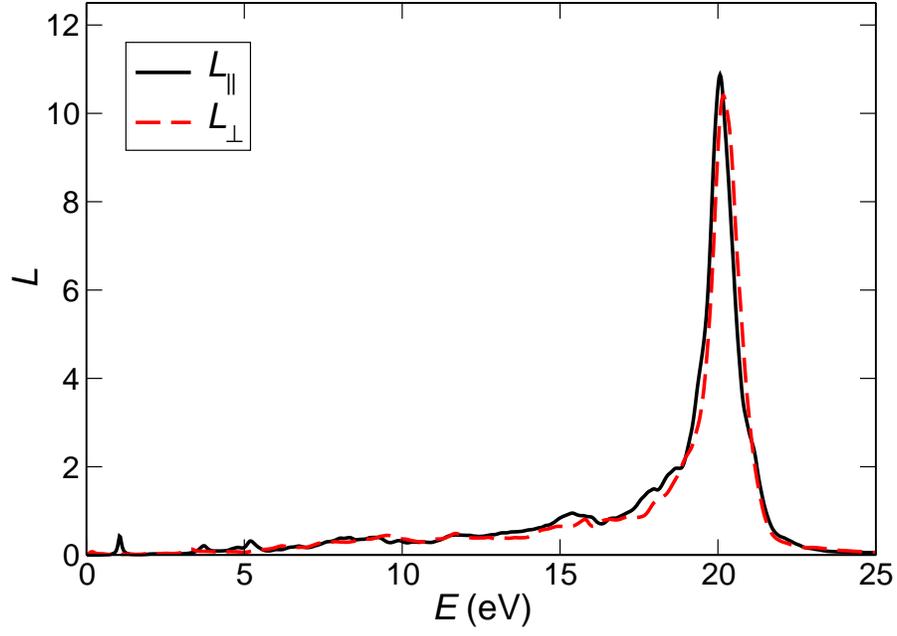}
   \caption{\label{eloss}
    Electron energy loss function $L$ has a pronounced peak close to the plasma
    frequency $\omega_p \approx 20$~eV.
    }
\end{figure}

Finally, the electron energy loss function 
$L_{ii}(\omega) = -\textrm{Im}\left(1/\varepsilon_{ii}(\omega)\right)$ is shown in 
figure~\ref{eloss}.
The pronounced peak at $\sim 20$ eV lies very close to $\omega_p$ and is 
intimately related to the plasma excitations;\cite{Pines} 
the effect of intraband transitions on this peak is small.

\section{Conclusions}
The aim of the paper is to stimulate synthesis and research on $\chem{Mo_6S_6}$
and $\chem{M_2Mo_6S_6}$ (M = alkali metal)
which - according to our theoretical predictions - are good quasi-one-dimensional
conductors and at the same time have high moduli of elasticity and resilience 
along the wires. This combination of electronic and mechanical properties 
makes them  unique in the area of one-dimensional
nanomaterials. The chains made of triangular $\chem{Mo_3}$, surrounded 
with three S, build conduction channels along the hexagonal $c$ axis with
very weak ``leakage'' of the current between individual chains. 
The intercalated alkali metals contribute additional
charge carriers and thus cause a shift in the Fermi energy but do not build 
their own conduction channels.

Synthesis of $\chem{Mo_6S_6}$ should follow the guidelines 
to make the $\chem{Mo_6Se_6}$\cite{Tarascon_85} 
whereas  $\chem{M_2Mo_6S_6}$ has been synthesized quite some time ago
together with  $\chem{M_2Mo_6Se_6}$.\cite{PCS80}
Synthesis might be difficult since in particular the compounds with S are 
metastable against the much more stable layered $\chem{MoS_2}$.
Intercalation with alkali metals, if necessary to stabilise the compound, 
does not change the described properties qualitatively.

$\chem{Mo_6S_6}$ and $\chem{M_2Mo_6S_6}$ could be challenging alternatives to 
other nanomaterials like carbon nanotubes.
The compounds have similar mechanical properties and thermal stability,
they are stable to about 700 K in the air (the thermal stability of 
$\chem{Mo_6S_6}$ is believed to be comparable to the stability of other
already existing Mo-S-I compounds). 
The advantage of $\chem{Mo_6S_6}$ and  $\chem{M_2Mo_6S_6}$ could be that they are  
easier to synthesise in a clean, one-stoichiometry form and that they have better 
conductivity than carbon nanotubes, due to their purely molybdenum metallic chains. 
$\chem{Mo_6S_6}$ is, according to the DFT results, also better conductor and
has better mechanical properties than the already synthesised Mo-S-I nanowires 
like $\chem{Mo_6S_{9-x}I_x}$. One dimensional nature of the systems opens also
the possibility of ballistic conduction or Peierls instability.

\begin{acknowledgments}
The author would like to thank D. Mihailovic for stimulating discussions. 
This work was  supported by the Slovenian Research Agency under
the contract  P1-0044. The crystal structures were visualized by 
Xcrysden.\cite{K99}
\end{acknowledgments}

  
\clearpage

\begin{table}
\caption{\label{latt} 
   The calculated lattice constants $(a,c)_\textrm{calc}$
   of  $\chem{Mo_6S_6}$ and  $\chem{K_2Mo_6S_6}$ 
   and the experimental lattice constants $(a,c)_\textrm{exp}$
   of $\chem{K_2Mo_6S_6}$.\cite{PCS80}}
\begin{tabular}[b]{ccc}
\hline
\hline
    &  $\chem{Mo_6S_6}$ & $\chem{K_2Mo_6S_6}$   \\
\hline
$a_\textrm{calc}$ &  $9.2\pm 0.1$ \AA  & $9.05\pm 0.1$ \AA \\
$a_\textrm{exp}$  &    ---             & 8.82 \AA \\
\hline
$c_\textrm{calc}$ &  $4.35\pm0.05$ \AA & $4.43\pm0.05$  \AA \\
$c_\textrm{exp}$  &    ---             &  4.44 \AA \\
\hline
\hline
\end{tabular}
\end{table}
\begin{table}
\caption{\label{dist}
Calculated ground-state interatomic distances in $\chem{Mo_6S_6}$. 
The in-plane distances are between the atoms of the same
Mo triangle with the adjacent three S atoms and the distances between 
the planes are between atoms in nearest neighbouring planes.
The numbers in the last column are for $c=4.63$ \AA, 
($\epsilon_{11} = 6.4$ \%), i.e., close to the point where the chains 
start to break.}

\begin{tabular}[b]{llll}
\hline
\hline
\multicolumn{2}{c}{Bond}  & \multicolumn{2}{c}{Distance} \\
                       &  & In equilibrium  & Strained   \\
\hline
Mo-S  & In-plane           &   2.49 \AA  & 2.46 \AA \\
Mo-S  & Between planes     &   2.62 \AA  & 2.65 \AA \\
Mo-Mo & In-plane           &   2.74 \AA  & 2.67 \AA \\
Mo-Mo & Between planes     &   2.69 \AA  & 2.79 \AA (2.782, 2.798)\\
S-S   & Between planes     &   3.60 \AA  & 3.66 \AA \\
S-S   & Closest interchain &   4.22 \AA  & 4.29 \AA \\
\hline
\hline
\end{tabular}

\end{table}
\begin{table}
\caption{\label{young_table}
Elastic properties of nanowires, carbon nanotubes (single-wall, SWCNT and multi-wall, MWCNT) and steel 
for comparison. $Y$ is the Young modulus, $E_R$ modulus of resilience, and $T$ the tensile strength.
The data for uniaxial systems are along the wires or tubes.
The ideal tensile strength of $\chem{Mo_6Se_6}$ is overestimated because it was assumed that the 
crystal stretches uniformly.}
\begin{minipage}[h]{\columnwidth}
\begin{tabular}[b]{lccc}
\hline
\hline
                  & $Y$ (GPa) &  $E_R$ (GPa) &  $T$ (GPa)        \\
\hline
$\chem{Mo_6S_6}$\footnote{Calculated in this paper} &  320      &   0.53       &  11  \\
$\chem{K_2Mo_6S_6}$$^a$ &  $\sim$230 &          &      \\
$\chem{Mo_6Se_6}$\footnote{Calculated using the data for an isolated molecular chain\cite{RRC02}} 
      &  250      &   3.1        &  26  \\
SWCNT\footnote{Compilation of data from Refs.~\onlinecite{YFAR00} and \onlinecite{OS03}}             
      & 1000      &     5 to 10     &  $\sim$30   \\
MWCNT                          & $< 1800$     &              &                 \\
Steel                          &   210     &   $\sim 3\times 10^{-3}$  &  0.4  \\
\hline
\hline
\end{tabular}
\end{minipage}
\end{table}
\clearpage

\begin{thebibliography}{xx}
\bibitem{PCS80}
  M. Potel, R. Chevrel, M. Sergent, J.C. Armici, M. Decroux, and \/O. Fischer,
  J. Solid State Chem. \textbf{35}, 286 (1980).
\bibitem{Tarascon_85} 
  J.M. Tarascon, J. Electrochem. Soc.  \textbf{132}, 2089 (1985).
\bibitem{VL99}
  L. Venkataraman and C.M. Lieber, Phys. Rev. Lett.  \textbf{83}, 5334 (1999).
\bibitem{BMPGS88}
  R. Brusetti, P. Monceau, M. Potel, P. Gougeon and M. Sergent, Solid State
  Comm.  \textbf{66}, 181 (1988).
\bibitem{RRC02}
  F.J. Ribeiro, D.J. Roundy and M.L. Cohen, Phys.Rev. B  \textbf{65}, 153401 
  (2002).
\bibitem{MKPG05}
A. Meden, A. Kodre, J. Padeznik Gomilsek, I. Arcon, I. Vilfan, D. Vrbanic,
A. Mrzel and D. Mihailovic, Nanotechnology  \textbf{16}, 1578 (2005).
\bibitem{BSMKL01}
  P. Blaha, K. Schwarz, G.K.H. Madsen, D. Kvasnicka, and J. Luitz, WIEN2k, 
  An APW + LO Program for Calculating Crystal Properties  
  (K. Schwarz, TU Vienna, 2001) ISBN 3-9601031-1-2.
\bibitem{SNS00} E. Sj\"{o}stedt, L. Nordstr\"{o}m, and D.J. Singh, Solid State
   Comm. \textbf{114}, 15 (2000).
\bibitem{PBE96}
  J.P. Perdew, K. Burke, and M. Ernzerhof, Phys. Rev. Lett. \textbf{77}, 3865
  (1996). 
\bibitem{KMRM03}
  A. Kis et al., Advanced Materials \textbf{15}, 733 (2003).
\bibitem{YFAR00}
  M.-F. Yu, B.S. Files, S. Arepalli and R.S. Ruoff, Phys. Rev. Lett. \textbf{84}, 5552 (2000).
\bibitem{OS03}
   S. Ogata and Y. Shibutani, Phys. Rev. B \textbf{68}, 165409 (2003).
\bibitem{CAD}
  C. Ambrosch-Draxl and J.O. Sofo, arXiv:cond-mat/0402523 (2004).
\bibitem{Pines}
  D.~Pines, Rev.~Mod.~Phys., \textbf{28}, 184 (1956).
\bibitem{K99}
       A. Kokalj: J. Mol. Graphics Modelling,  \textbf{17},
       176 (1999). Code available from http://www.xcrysden.org/.
\end{thebibliography}
\end{document}